\documentclass[epsfig,12pt]{article}
\usepackage{amsmath, amsthm, amssymb,slashed,graphicx,epsfig}

\newcommand {\slsh} [1] {\not{\hbox{\kern-2pt${#1}$}}}

\newcommand{\gsim}{\lower.7ex\hbox{$\;\stackrel{\textstyle>}{\sim}\;$}}
\newcommand{\lsim}{\lower.7ex\hbox{$\;\stackrel{\textstyle<}{\sim}\;$}}
\newcommand {\beq} {\begin{equation}}
\newcommand {\eeq} {\end{equation}}
\newcommand {\beqn}{\begin{eqnarray}}
\newcommand {\eeqn} {\end{eqnarray}}
\newcommand{\bea}{\begin{eqnarray}}
\newcommand{\eea}{\end{eqnarray}}

\begin{document}

%
\begin{titlepage}
\begin{center}
{\Large \bf William I. Fine Theoretical Physics Institute \\
University of Minnesota \\}
\end{center}
\vspace{0.2in}
\begin{flushright}
FTPI-MINN-12/27 \\
UMN-TH-3115/12 \\
August 2012 \\
\end{flushright}
\vspace{0.3in}
\begin{center}
{\Large \bf A Remark on Supersymmetric Bubbles and Spectrum Crossover
\\}
\vspace{0.2in}
{\large M.~Shifman$^{a,b}$  and M.B. Voloshin$^{a,b,c}$  \\ [0.2in] }
$^a$School of Physics and Astronomy, University of Minnesota, Minneapolis, MN 55455, USA \\
$^b$William I. Fine Theoretical Physics Institute, University of
Minnesota,\\ Minneapolis, MN 55455, USA \\
$^c$Institute of Theoretical and Experimental Physics, Moscow, 117218, Russia
\\[0.2in]

\end{center}

\vspace{0.2in}

\begin{abstract}
Using an exact expression for the domain wall tension
in a supersymmetric model we show that
a spectrum crossover takes place in passing from weak to strong coupling.
In the weak coupling  regime
 elementary excitations are the lightest states, while in the strong coupling regime
  solitonic objects of a special type -- bubbles -- assume the role of the lightest states.
  The crossover occurs at $ {\lambda^2}/(4\pi) \sim 0.4.$
\end{abstract}
\end{titlepage}

The (possible) recent discovery of the Higgs particle, with the production and decay properties fully consistent with the standard model (SM) implies that the scale of new physics is higher than we hoped. Of special importance is the $\gamma\gamma$ decay of the Higgs particle which agrees, within errors, with the SM prediction \cite{one}. 
The theoretical number \cite{one} is practically impossible to change without drastic modifications of the electroweak theory.\footnote{An example of a dramatic crippling of the theory needed
to enhance $\Gamma (H\to \gamma\gamma )$ just by a factor of 2
is presented in \cite{ArkaniHamed:2012kq}.} 
 Not only this agreement is remarkable, but we  learn, 
from the fact that $m_H \sim 125\,$GeV, 
 that the theory, while  keeping itself at weak coupling, comes rather close to the boundary of 
the weak coupling regime, since the Higgs self-interaction coupling $\lambda \sim \frac12$.  

It is not ruled out that future beyond-SM explorations will uncover a more complicated Higgs sector, 
with still larger coupling constants. In the strong coupling theories the phenomenon of level crossing
is quite common. There is a problem with its detection, because usually it occurs at strong coupling. We are aware of several examples:
(i) in two-dimensional models with exact solutions \cite{Coleman:1974bu}, where 
at weak coupling the lightest state is an elementary excitation, while at strong coupling it is a soliton;
 (ii) in supersymmetric theories in two and four dimensions in the BPS protected sectors,
the so-called curves of marginal stability or domain wall crossings
 (where this knowledge is essentially algebraic, plus analytic properties) \cite{four}; (iii) 
in supersymmetric theories with dualities, in the non-BPS sectors, the so-called crossover \cite{five}. 

In the latter case, the dynamical information needed to detect the level crossing is provided by a
 weak-strong coupling duality. Here we discuss a simple example in which the necessary dynamical information
 comes from some general considerations combining supersymmetry and quantum mechanics.
 The dynamical systems that we keep in mind, that become light at strong coupling 
 are bubbles of an ``opposite" vacuum.
 
 Such bubbles were considered in the literature previously, on several occasions \cite{six,seven}.
 In \cite{six} highly excited bubble states were considered at weak coupling, where they are much heavier 
 than the elementary excitations of the model. High excitation number was crucial 
 for maintaining a well defined bubble (albeit unstable). In \cite{seven} pure supersymmetric $\mathcal{N}=1$
 Yang-Mills theory was treated. Needless to say, this is a strongly coupled theory with $N$ 
 degenerate vacua (if the gauge group is SU$(N)$). However, the domain wall tension $T$
  scales as $T\sim N$ \cite{eight}, hence $T^{1/3}$ is always larger than the glueball mass, and therefore the bubbles 
 under consideration
  are unstable. Strictly speaking they are absent in the spectrum of the stable states.\footnote{The authors of \cite{seven}
  understand the bubble instability; however, they argue that the bubbles are quasistable at large $N$. 
  Their argument does not seem persuasive at all given the fact that a typical bubble size scales as 
  $T^{-1/3} \ll \Lambda^{-1}$. It is true, though, that the bubbles in \cite{seven} are in the thin wall regime, 
  see \cite{nine} for an explanation.
  }
  In both cases \cite{six,seven}, the crossover phenomenon is untraceable. 
 
 We will focus on a simple supersymmetric set-up with the weak-strong coupling transition.
 We  consider a minimal ${\mathcal N}=1$ Wess-Zumino model with the superpotential
\beq
 {\mathcal W} = \frac{m^2}{\lambda}\Phi - \frac{\lambda}{3}\Phi^3\,,
 \label{one}
\eeq
assuming for simplicity the mass and $\lambda$ parameters to be real and positive.

\begin{figure}[h]
\epsfxsize=7cm
\centerline{\epsfbox{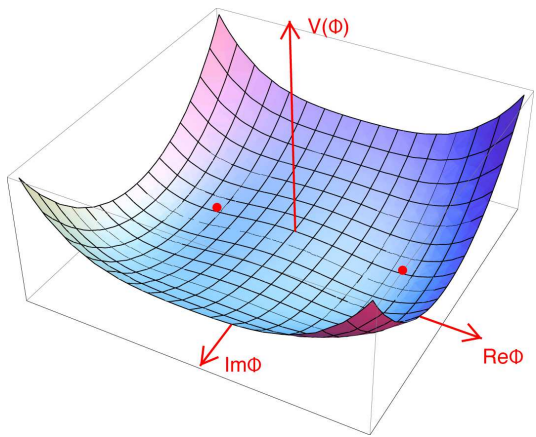}}
\caption{\small Potential energy in the model (\ref{one}). Two degenerate isolated minima are marked by dots.}
\label{pot}
\end{figure}

 This theory has two isolated degenerate supersymmetric vacua at 
 \beq
 \phi_{\rm vac} = \pm \frac{m}{\lambda}\,,
 \eeq
and a domain wall interpolating between them. The domain wall is BPS saturated and, therefore, its 
tension $T$ is exactly known (see e.g. \cite{ten}),
\beq
T= \frac{8}{3} \frac{m^3}{\lambda^2}\,.
\label{three}
\eeq
Note that the combination appearing on the right-hand side is renormalization group invariant.
We can write this ratio either in terms of the bare $m_0$ and $\lambda_0$, or in terms of
 the physical mass $m$ and coupling constant $\lambda$, the result is the same. 
 
 At weak coupling, $\lambda \ll 1$, the parameter $m$ is the mass of the elementary excitation
 in either of the two vacua. This is the lightest (and the only bosonic) particle in the theory.
 We claim that if we fix $m$ and analytically continue $\lambda$ to strong coupling, the would-be ``elementary" 
 excitations decay. A number of lighter stable states of a totally different nature
 appear in the theory.

These states are bubbles of the ``opposite" vacuum, analogous to those occurring in 
the problem of the false vacuum decay~\cite{KOB,Coleman}. In the latter problem the decaying vacuum has a higher energy density
than the genuine vacuum. Therefore, one has to deal with the volume energy of the bubble.
In the supersymmetric case at hand both vacua are degenerate. As a result the bubble dynamics is determined by that
of its surface. This dynamics can be readily described as a quantum-mechanical problem in the
so-called thin wall approximation.

The wall thickness is determined by $m^{-1}$ while its radius, as we will see shortly, is proportional to
$T^{-1/3}$. In this approximation one can neglect the deformation of the
wall tension due to a nonvanishing curvature. Therefore, we need $m\gg T^{1/3}$, which, implies, in turn, strong coupling.
This domain is amenable to studies due to the exact nature of Eq. (\ref{three}).
Assuming the bubble to be spherical one can write a quantum-mechanical Lagrangian governing its
dynamics,
\beq
{\mathcal L} = -4\pi R^2 T \sqrt{1- \dot{R}^2}\,,
\label{effl}
\eeq
where $R$ is the bubble radius. The corresponding relation for the Hamiltonian ${\mathcal H}$ in terms of $R$ and the conjugate momentum $p$ reads as
\beq
{\mathcal H}^2 -p^2  = \left ( 4 \pi R^2 T \right )^2 \, .
\label{ham}
\eeq
The ground state energy $E_0$ of a quantized bubble described by this Hamiltonian can be readily found by numerically solving the Schr\"odinger equation corresponding to Eq.(\ref{ham}),
\beq
E_0 = c_0 \, (4 \pi)^{1/3} \, T^{1/3} \approx  3.32 \, \frac{m}{\lambda^{2/3}}\,,
\eeq
with $c_0=1.027 \ldots$. 

Generically, the energy of the $n$-th quantized state of the bubble can be written as $E_n = c_n \, (4 \pi)^{1/3} \, T^{1/3}$ with $c_n$ being a dimensionless coefficient. For the first excited bubble the numerical solution gives 
$c_1=1.949 \ldots $ and this state is stable with respect to decay into two ground-state bubbles, since $c_1 < 2 \, c_0$. 

At large $n$ for
spherically symmetric bubbles the coefficients $c_n$ can be evaluated by using the Bohr-Sommerfeld type semiclassical quantization applied to Eq. (\ref{ham}). In this way one finds
\beq
c_n=\left [ {3 \sqrt{\pi} \, \Gamma(3/4) \over  \Gamma(1/4)} \right]^{1/3} \, \left ( n + {1 \over 2} \right )^{2/3}\, ~~~~ (n \gg 1)\,.
\label{boso}
\eeq
None of the excited states with $n > 1$ is stable with respect to decay into ground-state bubbles.
(It can be noted that although the semiclassical expression (\ref{boso}) is justified at large $n$, formally setting $n=0$ and $n=1$, one finds that it gives $c_0 \approx 0.93$ and $c_1 \approx 1.937$; these values only slightly differ from the exact ones. Thus, the semiclassical formula works reasonably well starting from low $n$.)

The condition for stability of the ground-state bubble against decay into two elementary bosons, $E_0 <2m$, implies
\beq
\lambda^{2/3} > 1.66 \quad {\rm or}\quad \alpha\equiv \frac{\lambda^2}{4\pi} > 0.365\,.
\label{seven}
\eeq
It should be noted however, that the specific numerical value in the above estimate should be taken with a certain reservation. Indeed, at such value of $\lambda$ 
we have $T^{-1/3} \gsim m^{-1}$; 
the condition $T^{-1/3} \gg m^{-1}$ is not met, and literally speaking 
the bubble cannot be considered in the thin wall approximation, i.e. by virtue  of the effective Lagrangian (\ref{effl}). 
At this point the thin wall approximation is at the boundary of its applicability.
It is  clear, however, that the ground-state bubble becomes stable and in fact the lowest mass state at a sufficiently 
large $\lambda$, somewhere 
above the limit (\ref{seven}). 

Of course, if the physical coupling constant satisfies Eq. (\ref{seven})
we are not that far from the Landau pole. To make the model under consideration
self-consistent one must assume that it has some ultraviolet (UV) completion that embeds it into an asymptotically free field theory
(or a non-field-theoretic UV completion).

If the condition (\ref{seven}) is met, the bubbles of the ``opposite" vacuum 
and, perhaps, a number of excitations form a spectrum of stable bosons (supersymmetry implies that there are degenerate
in mass  fermions too). Needless to say, their masses are not BPS protected. Thus, we deal here with long supersymmetry multiplets.
A typical size of the above states is determined by $T^{-1/3}$ rather than by $m^{-1}$. 

The very idea of building various solitonic objects by bending domain walls and stabilizing them 
appropriately, is not new, of course. We have already mentioned \cite{six, seven}. In addition, in \cite{bol}
magnetic flux tubes were constructed in this way in supersymmetric non-Abelian Yang-Mills theories.
The peculiarity of the example we have considered in this note is that
by varying the value of the coupling constant $\lambda$ 
we can travel all the way from the weak coupling  regime
in which the elementary excitations are the lightest states, to the strong coupling regime
in which solitonic objects of a special type -- bubbles -- assume the role of the lightest states.

In conclusion we note that, although nothing can be proven without supersymmetry, it is not ruled out that
a similar phenomenon occurs in  nonsupersymmetric models with spontanepusly broken discrete symmetries
in passing from weak to strong coupling.

	\section*{Acknowledgments}

\noindent

This work was supported by the DOE grant DE-FG02-94ER40823.

\end{document}